\documentclass[a4paper,11pt]{article}
\usepackage{graphicx}

\begin{document}


\title{Continuous variable teleportation of \\ single photon states}

\author{Toshiki Ide$^a$, Holger F. Hofmann$^b$, Takayoshi Kobayashi$^a$ \\
and Akira Furusawa$^c$ \\
$^a$Department of Physics, Faculty of Science, University of Tokyo,\\
7-3-1 Hongo, Bunkyo-ku, Tokyo113-0033, Japan \\
$^b$CREST, Japan Science and Technology Corporation (JST),\\
Research Institute for Electronic Science, Hokkaido University, \\
Sapporo 060-0812, Japan \\
$^c$Department of Applied Physics, \\ Faculty of Engineering, University of Tokyo,\\
7-3-1 Hongo, Bunkyo-ku, Tokyo113-8656, Japan \\
}

\date{}

\maketitle

\begin{abstract}
The properties of continuous variable teleportation of single photon states are investigated. The output state is different from the input state due to the non-maximal entanglement in the EPR beams. 
The photon statistics of the teleportation output are determined and the correlation between the field information $\beta$ obtained in the teleportation process and the change in photon number is discussed. The results of the output photon statistics are applied to the transmission of a qbit encoded in the polarization of a single photon.  
\end{abstract}

\section{Introduction}

Quantum teleportation is a method for Alice (sender) to transmit an unknown quantum input state to Bob (receiver) at a distant place by sending only classical information using a shared entangled state as a resource.
Originally quantum teleportation was proposed for discrete variables in two-dimensional Hilbert spaces\cite{Ben93}.
Later it was applied to continuous variables (two components of the electromagnetic field) in infinite-dimensional Hilbert spaces\cite{Vai94}.
However, continuous variable teleportation ideally
 requires a maximally entangled state which has infinite energy as a resource. 
Nevertheless it has been 
shown theoretically that quantum teleportation realized by using non-maximally entangled state can still transfer non-classical features of quantum states \cite{Brau98}.
Experimentally, such a continuous variable teleportation has been realized by Furusawa {\it et.al.} \cite{Fur98}. In \cite{Brau98}, the physics of continuous variable teleportation was described in terms of Wigner function. Ref.\cite{Enk99} described it in terms of discrete basis states. 
Ref.\cite{Hof00} formulates the whole process of the quantum teleportation by a transfer operator which is acting on arbitrary input states.

In the experiment of \cite{Fur98}, a coherent state was used as an input state. But any quantum state can be teleported by this method. Therefore, the transfer of non-classical states is of interest.
In the following, the transfer operator formalism derived in \cite{Hof00} is used for analyzing the photon statistics of the output state of a one photon state teleportation.
It is shown that the change in photon number is strongly dependent on the field measurement result obtained in the process of teleportation. This result is then applied to the two mode teleportation of
a polarized photon, illustrating the possibility of using continuous
variable teleportation for the transfer of single photon qbits.

\section{Transfer operator}

\begin{figure}
\begin{picture}(400,300)

\put(160,40){\framebox(80,40){\Large OPA}}

\put(160,85){\line(-1,1){45}}
\put(155,80){\line(-1,1){45}}
\put(110,130){\line(0,-1){10}}
\put(110,130){\line(1,0){10}}
\put(135,105){\makebox(20,20){\Large R}}

\put(240,85){\line(1,1){45}}
\put(245,80){\line(1,1){45}}
\put(290,130){\line(0,-1){10}}
\put(290,130){\line(-1,0){10}}
\put(245,105){\makebox(20,20){\Large B}}

\put(40,85){\line(1,1){45}}
\put(45,80){\line(1,1){45}}
\put(90,130){\line(0,-1){10}}
\put(90,130){\line(-1,0){10}}
\put(45,105){\makebox(20,20){\Large A}}
\put(10,66){\makebox(40,12){\large Input}}
\put(10,54){\makebox(40,12){\large field}}

\put(100,110){\line(0,1){60}}
\put(80,90){\makebox(40,12){\large Beam}} 
\put(80,78){\makebox(40,12){\large splitter}}

\put(90,155){\line(-1,1){45}}
\put(85,150){\line(-1,1){45}}
\put(40,200){\line(0,-1){10}}
\put(40,200){\line(1,0){10}}
\put(30,200){\makebox(20,20){\Large $x_-$}}

\put(110,155){\line(1,1){45}}
\put(115,150){\line(1,1){45}}
\put(160,200){\line(0,-1){10}}
\put(160,200){\line(-1,0){10}}
\put(150,200){\makebox(20,20){\Large $y_+$}}

\put(40,230){\framebox(120,60){}}
\put(60,260){\makebox(80,20){\large Measurement of}}
\put(60,240){\makebox(80,20){\large $\beta=x_-+i y_+$}}

\bezier{400}(160,260)(250,260)(290,180)
\put(290,180){\line(0,1){12}}
\put(290,180){\line(-3,2){10}}

\put(295,135){\framebox(40,40){\Large $\hat{D}(\beta)$}}

\put(340,185){\line(1,1){25}}
\put(345,180){\line(1,1){25}}
\put(370,210){\line(0,-1){10}}
\put(370,210){\line(-1,0){10}}

\put(350,227){\makebox(40,12){\large Output}}
\put(350,215){\makebox(40,12){\large field}}
\end{picture}
\caption{\label{setup} Schematic representation of the quantum teleportation 
setup.}
\end{figure}
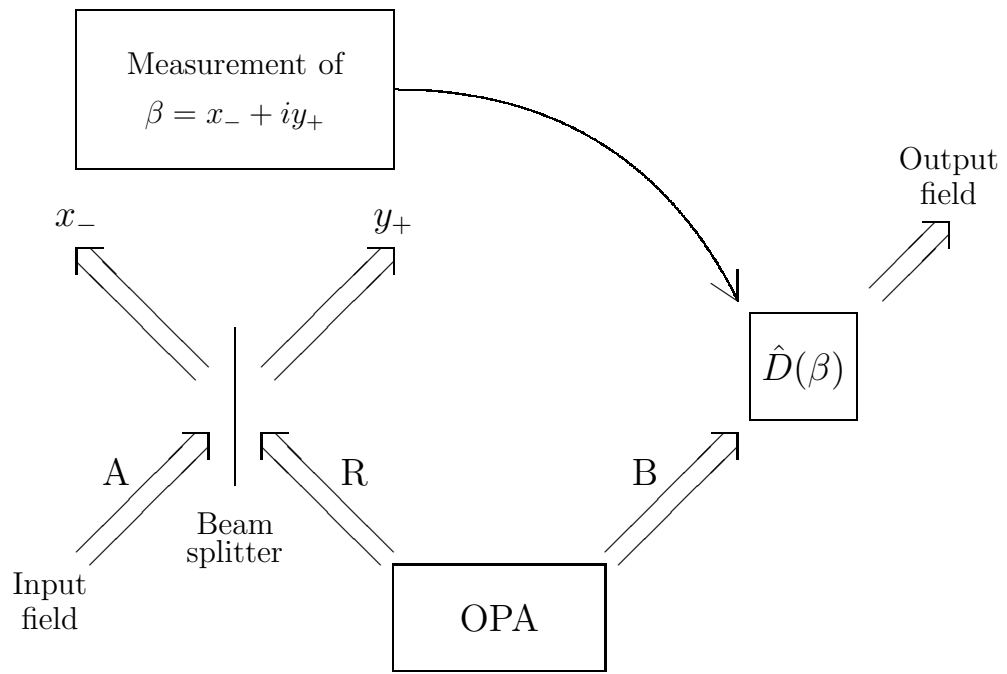

Fig.\ref{setup} shows the schematic sets of the quantum teleportation according to \cite{Fur98}.
Alice transmits an unknown quantum state $\mid \psi \rangle _{A}$ to Bob. Alice and Bob share EPR beams in advance. 
The quantum state of the EPR beams reads \cite{Enk99,Hof00}
\begin{equation}
\mid q \rangle_{R,B} = \sqrt{1-q^2} \sum_{n=0}^{\infty} q^n \mid n \rangle_{R} \mid n \rangle_{B}.
\end{equation}
where R,B are the modes for reference and Bob each. $q$ is a parameter which stands for the degree of entanglement. It varies from 0 to 1 with 1 being maximal entanglement and 0 being no entanglement. The degree of entanglement depends on the squeezing achieved in the parametric amplification. 
In the experiment of \cite{Fur98}, 3dB ($q=0.33$) squeezed light was used. 
Squeezing of up to 10dB ($q=0.82$) should be possible with available technology. 

Alice mixes her input state with the reference EPR beam by a 50$\%$ beam-splitter and performs an entanglement measurement of the complex field value $\beta = x_{-}+iy_{+}$, where
\begin{eqnarray}
\hat{x}_{-}&=&\hat{x}_{A}-\hat{x}_{R}, \nonumber \\
\hat{y}_{+}&=&\hat{y}_{A}+\hat{y}_{R}.
\end{eqnarray}
This measurement projects $A$ and $R$ onto the eigenstate
\begin{equation}
\mid \beta \rangle_{A,R} = \frac{1}{\sqrt{\pi}}\sum_{n=0}^{\infty}
\hat{D}_{A}(\beta)
\mid n \rangle_{A} \mid n \rangle_{R},
\end{equation}
where $\hat{D}_{A}(\beta)$ is a displacement operator acting on the mode A 
with a displacement amplitude of $\beta$.
The output state $\mid \psi(\beta)\rangle_{B}$ conditioned by the measurement process can be written as a projection of the initial product state onto the eigenstate  $\mid \beta \rangle_{A,R}$, 
\begin{eqnarray}
\mid \psi(\beta)\rangle_{B}&=&_{A,R}\langle\beta\mid\psi\rangle_{A}\mid q \rangle_{R,B} \nonumber \\
&=& \sqrt{\frac{1-q^2}{\pi}}\sum_{n=0}^{\infty}q^n \mid n \rangle_{BA} \langle n \mid \hat{D}_{A}(-\beta)\mid \psi\rangle_{A} .
\end{eqnarray}
$\mid \psi(\beta)\rangle_{B}$ is not normalized since the probability of 
obtaining the field measurement value $\beta$ is given by $P_{q}(\beta)=\langle \psi_{\mbox{out}}(\beta) \mid \psi_{\mbox{out}}(\beta) \rangle $.

After Bob gets the information of the field measurement value $\beta$ from Alice, Bob applies a displacement to the output state by mixing the coherent field of a local oscillator with the output EPR beam $B$. The output state $\mid~\!\!\psi_{\mbox{out}}(\beta)\rangle_{B}=\hat{D}_{B}(\beta)\mid\psi (\beta)\rangle_{B}$ 
may also be written as
\begin{equation}
\mid \psi_{\mbox{out}}(\beta) \rangle_{B} = \hat{T}_{q}(\beta) \mid \psi\rangle_{A},
\end{equation}
where $\hat{T}_{q}(\beta)$ is a transfer operator which represents all processes of the quantum teleportation \cite{Hof00}. In its diagonalized form, it reads
\begin{equation}
\hat{T}_{q}(\beta) = \sqrt{\frac{1-q^2}{\pi}}\sum_{n=0}^{\infty}q^n \hat{D}(\beta) \mid n \rangle _{BA}\langle n \mid \hat{D}(-\beta).
\label{transfer}
\end{equation}
When $q\rightarrow 1$, $\hat{T}_{q}(\beta)$ becomes similar to the equivalence operator $\bf{\hat{1}}$ except for a change of the mode from A to B  which indicates a perfect teleportation with no modification to the input state. 
For general $q$, the transfer operator at $\beta=0$ is diagonal in the photon number states.

\section{One photon state teleportation}

In the experiment of \cite{Fur98}, a coherent state was transfered. In this case, the output state is modified but is still a coherent state \cite{Hof00}. 
In case of a one photon input state , the output state is 
\begin{eqnarray}
\hat{T}_{q}(\beta) \mid 1 \rangle 
&=& 
\sqrt{\frac{1-q^2}{\pi}}\sum_{n=0}^{\infty}q^n \hat{D}(\beta) \mid n \rangle \langle n \mid \hat{D}(-\beta)
\mid 1 \rangle \nonumber\\
&=& 
\sqrt{\frac{1-q^2}{\pi}}\sum_{n=0}^{\infty}q^n \hat{D}(\beta) \mid n \rangle \langle n \mid \hat{D}(-\beta)\hat{a}^{\dagger}
\mid 0 \rangle \nonumber\\
&=& 
\sqrt{\frac{1-q^2}{\pi}}\sum_{n=0}^{\infty}q^n \hat{D}(\beta) \mid n \rangle \langle n \mid (\hat{a}^{\dagger}+\beta^{\ast})\hat{D}(-\beta)
\mid 0 \rangle \nonumber\\
&=& 
\sqrt{\frac{1-q^2}{\pi}} e^{-(1-q^2) \frac{|\beta|^2}{2}} 
\hat{D}((1-q)\beta)
\left ( (1-q^2) \beta^{\ast} \mid 0 \rangle + q \mid 1 \rangle \right ). \nonumber \\
\ 
\label{out}
\end{eqnarray}

For simplicity, the suffixes for the modes are not explicitly given. 
In general, this output state is quite different from the original one photon input state.

$\beta$ represents a coherent field measurement performed on the input state.
The probability distribution over field measurement values $\beta$ for one photon input states is given by 
\begin{eqnarray}
P_{q}(\beta)
&=&\langle \psi_{\mbox{out}}(\beta) \mid \psi_{\mbox{out}}(\beta) \rangle \nonumber \\
&=&\langle 1 \mid \hat{T}_{q}^{\dagger}(\beta) \hat{T}_{q}(\beta) \mid 1 \rangle \nonumber \\
&=&\frac{1-q^2}{\pi}e^{-(1-q^2)|\beta|^2}
\left ( (1-q^2)^2 |\beta|^2+ q^2 \right ).
\end{eqnarray}

\begin{figure}[ht]
\begin{picture}(350,400)
\put(40,150){\makebox(70,30){\Large $P_{\frac{1}{2}}(\beta)$}}
\put(40,0){\makebox(300,200){\includegraphics[width=10cm]{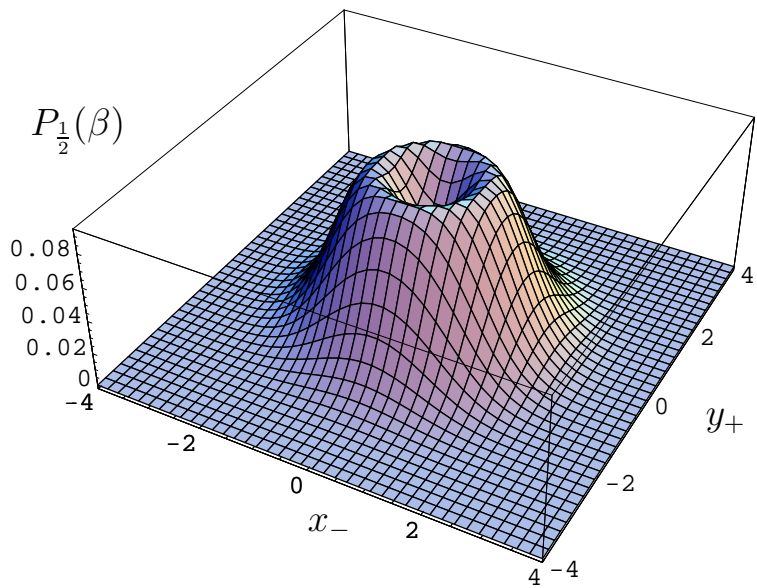}}}
\put(150,10){\makebox(40,10){\Large $x_{-}$}}
\put(300,50){\makebox(40,10){\Large $y_{+}$}}
\end{picture}
\caption{The probability distribution of $P_{q}(\beta)$ for the field measurement value $\beta=x_{-}+iy_{+}$ for $q=\frac{1}{2}$.}
\label{p_b}
\end{figure}

Fig.\ref{p_b} shows the probability distribution $P_{q}(\beta)$ for $q=\frac{1}{2}$.
The circular symmetric distribution with a dip in the middle and a maximum for amplitude of $|\beta|=1$ is  characteristic of the one photon input state. 

\begin{figure}[ht]
\begin{picture}(350,400)
\put(40,190){\makebox(70,20){\Large $P_{\frac{1}{2}}(n)$}}
\put(40,0){\makebox(300,200){\includegraphics[width=10cm]{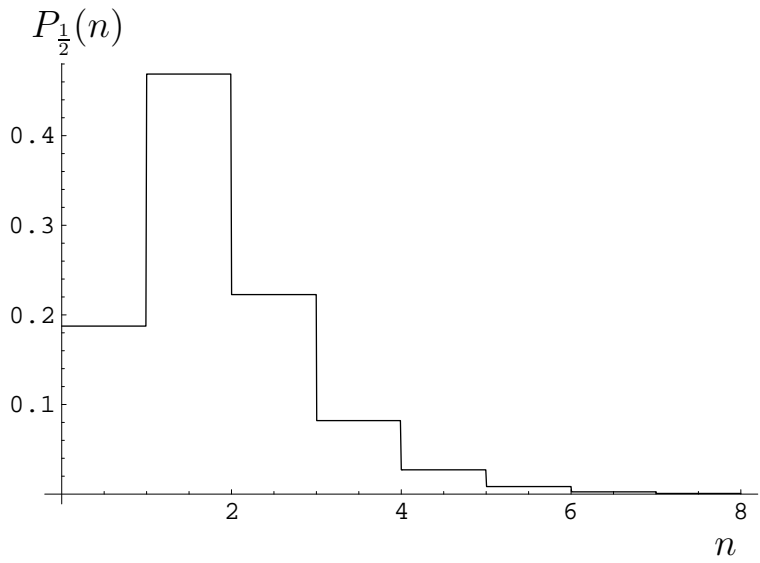}}}
\put(300,0){\makebox(40,10){\Large $n$}}
\end{picture}
\caption{The probability of $n$-photon counting $P_{q}(n)$ for one photon state teleportation in the case of $q=\frac{1}{2}$.}
\label{p_n}
\end{figure}

The properties of the output state can be investigated experimentally by several kinds of detection setups. 
If Bob has a photon counting setup which can discriminate photon numbers, he can obtain the photon statistics of the output state.
The overall photon statistics of the output are obtained by integrating over $\beta$, 
\begin{eqnarray}
P_{q}(n)&=&\int d^2\beta | \langle n \mid \hat{T}_{q}(\beta)\mid 1 \rangle|^2 \nonumber \\
&=& \frac{1+q}{2} \left ( \frac{1-q}{2} \right )^{n+1} 
\left ( 1+ \left ( \frac{1+q}{1-q} \right )^2 n \right ).
\label{Pn}
\end{eqnarray}
$P_q(n)$ is the probability of counting $n$-photons after the teleportation. Fig.\ref{p_n} shows the probability distribution over $n$ in the case of $q=\frac{1}{2}$.
The probability of detecting one photon is maximal and 
the probabilities of higher photon numbers get less likely as photon number increases.

\begin{figure}[ht]
\begin{picture}(350,400)
\put(40,190){\makebox(70,20){\Large $P_{q}(n)$}}
\put(40,0){\makebox(300,200){\includegraphics[width=10cm]{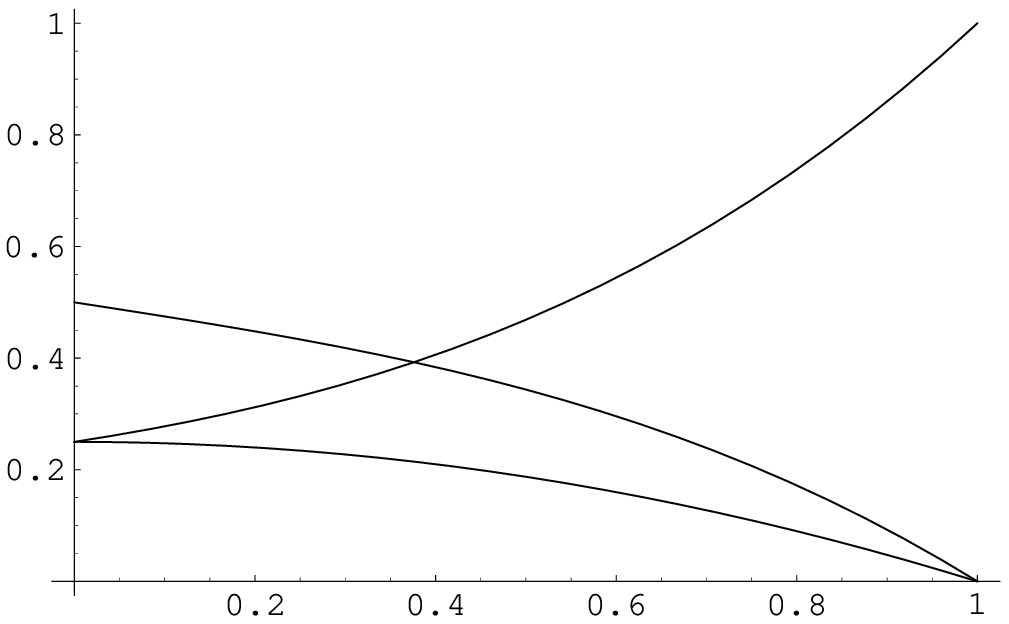}}}
\put(250,100){\makebox(40,10){\Large $P_{q}(1)$}}
\put(100,40){\makebox(40,10){\Large $P_{q}(0)$}}
\put(300,50){\makebox(40,10){\Large $P_{q}(n\geq 2)$}}
\put(300,0){\makebox(40,10){\Large $q$}}
\end{picture}
\caption{The probability of zero photon, one photon and more than one photon counting for one photon state teleportation.}
\label{p_012}
\end{figure}

The changes in photon number may be summarized in terms of photon loss $n=0$,
successful photon transfer $n=1$, and photon gain $n\geq2$. The $q$ dependence of these probabilities is given by 
\begin{eqnarray}
P_{q}(0)&=&\frac{1}{4}(1-q^2), \\
P_{q}(1)&=&\frac{1}{4}(1+q+q^2+q^3), \label{fidelity} \\
P_{q}(n\geq 2)&=&\frac{1}{4}(2-q-q^3).
\end{eqnarray}
Since the input state is a one photon state, eq. (\ref{fidelity}) gives the fidelity of the teleportation. 
Fig.\ref{p_012} shows the $q$ dependence of $P_{q}(0),P_{q}(1),P_{q}(n\geq 2)$.
In the case of a maximally entangled state ($q\rightarrow$1), Bob can receive nothing but a one photon state ($P_{q}(1)\rightarrow$1) and all other probabilities vanish ($P_{q}(0)\rightarrow$0,$P_{q}(n\geq 2)\rightarrow$0), which indicates perfect teleportation.
In the case of non-maximally entangled states $(q<1)$, the probabilities of zero-photon counting and more than one-photon counting appear. 
Note that the probability of photon gain $n\geq 2$ is always greater  than that of photon loss $n=0$. 

In order to investigate the change in the photon number given by eq.(\ref{Pn}) in more detail, we now derive the conditional probability distributions over $\beta$ for the cases of zero-photon, one-photon, and more than one-photon counting.
These probabilities can be obtained from $P_{q}(n,\beta)= |\langle n \mid \hat{T}_{q}(\beta)\mid 1 \rangle|^2$. The results read 
\begin{eqnarray}
P_{q}(0,\beta)&=&\frac{1-q^2}{\pi} e^{-2(1-q)|\beta|^2} (1-q)^2 |\beta|^2 , \\
P_{q}(1,\beta)&=&\frac{1-q^2}{\pi} e^{-2(1-q)|\beta|^2} 
\left ( (1-q^2)^2 |\beta|^2 +q^2 \right ) , \\
P_{q}(n\geq 2,\beta)&=&P_{q}(\beta)-P_{q}(0,\beta)-P_{q}(1,\beta).
\end{eqnarray}

\begin{figure}[ht]
\begin{picture}(350,400)
\put(40,190){\makebox(270,20){\Large $P_{\frac{1}{2}}(\beta)$}}
\put(40,0){\makebox(300,200){\includegraphics[width=10cm]{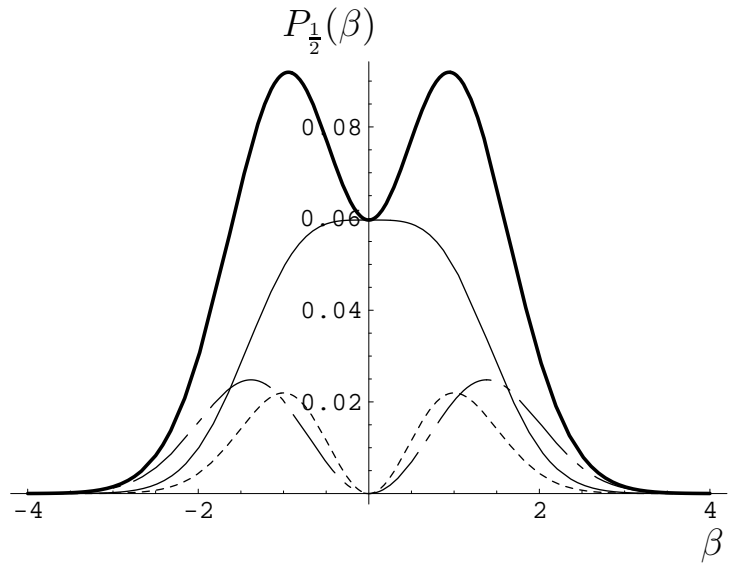}}}
\put(300,0){\makebox(40,10){\Large $\beta$}}
\end{picture}
\caption{The probability of $n$-photon counting $P_{q}(n)$ for one photon state teleportation in the case of $q=\frac{1}{2}$. 
The thick solid line shows $P_{\frac{1}{2}}(\beta)$, the thin solid line shows $P_{\frac{1}{2}}(1,\beta)$, the dotted line shows $P_{\frac{1}{2}}(0,\beta)$ and the dot-bar line shows $P_{\frac{1}{2}}(n\geq2,\beta)$.
}
\label{p_tot}
\end{figure}

Fig.\ref{p_tot} shows these different contributions to the total probability distribution over $|\beta|$ in case of $q$=0.5. 
For high values of $|\beta|$, every probability vanishes.
In the $\beta=0$ case, we see that only the probability of obtaining a one photon output is non-zero. This means that the output of the EPR beam $B$ is already a one photon state without any additional displacement. 
In general,  
 the photon number states are the eigenstates of $\hat{T}_{q}(\beta=0)$, as can be seen from eq.(\ref{transfer}). A measurement value of $\beta=0$ thus indicates that the output beam photon number is equal to the input beam photon number. 
Note also that a low field amplitude provides little information about the phase of the teleported field.  
Therefore, the lack of phase information in the input photon number state is preserved in the teleportation process. 

In the region of $|\beta|\ll1$, the probability of obtaining a one photon output is nearly constant and that of zero-photon and more than one-photon are increasing with the increase of $|\beta|$. 
The increase of these two probabilities gives rise to the peak of the total probability. 
The probability of photon loss ($n=0$) and photon gain ($n\geq 2$) are crossing  at around $|\beta|=1$. 
Below $|\beta|=1$ the probability of photon loss ($n=0$) is greater than that of photon gain ($n\geq 2$). 
Above $|\beta|=1$, the probability of photon gain ($n\geq 2$) is dominant. 
For high field measurement results $|\beta|\gg 1$, 
the teleportation process generates more photons in the output.
Since $\beta$ can be considered as a measurement of the coherent input field amplitude, this result is similar to the correlation of field measurement and photon number discussed in \cite{HofQND}.

\section{Application to single photon polarization}
Since the polarization of the light field can be used to encode information using single photons, the case of a polarization sensitive continuous variable teleportation is of considerable interest. The results obtained for the single mode case can be applied to the teleportation of two polarization modes, H and V, by applying a separate transfer operator to each mode. 
Note that this does not imply that 
the teleportation $H$ and $V$ has to be conducted separately. 
For example, the two dimensional measurement amplitude $(\beta_H,\beta_V)$ could also be obtained by measuring the circular polarization components $(\beta_H \pm i \beta_V)$. The continuous variable 
teleportation of polarized photons therefore does not require any
previous knowledge of the input polarization and the results 
obtained for successful transfers and for polarization flips can be
applied directly to the teleportation of an unknown photon 
polarization. 

The output state of a one photon state with polarization $H$ can be 
written as a product of the one photon teleportation given in 
equation (\ref{out}) and a vacuum teleportation, which is a special
case of the conventional coherent state teleportation discussed
in \cite{Hof00}. The result reads  
\begin{eqnarray}
&&\hat{T}_{Hq}(\beta_H) \hat{T}_{Vq}(\beta_V) \mid 1 \rangle_H \mid 0 \rangle_V  \nonumber\\
&=& \hspace{1em}
\sqrt{\frac{1-q^2}{\pi}}\sum_{n=0}^{\infty}q^n \hat{D}_H(\beta_H) \mid n \rangle_{HH} \langle n \mid \hat{D}_H(-\beta_H)
\mid 1 \rangle_H \nonumber\\
&&\otimes 
\sqrt{\frac{1-q^2}{\pi}}\sum_{m=0}^{\infty}q^m \hat{D}_V(\beta_V) \mid m \rangle_{VV} \langle m \mid \hat{D}_V(-\beta_V)
\mid 0 \rangle_V \nonumber\\
&=& \hspace{1em}
\sqrt{\frac{1-q^2}{\pi}}
e^{-(1-q^2) \frac{|\beta_H|^2}{2}}
\hat{D}((1-q)\beta_H)
\left ( (1-q^2) \beta_H^{\ast} \mid 0 \rangle_H + q \mid 1 \rangle_H \right )
 \nonumber \\ 
&&\otimes \sqrt{\frac{1-q^2}{\pi}}
e^{-(1-q^2) \frac{|\beta_V|^2}{2}}
\hat{D}((1-q)\beta_V) \mid 0 \rangle_V .
\ 
\label{pout}
\end{eqnarray}
This result describes all details of single photon teleportation,
including the information obtained from the measurement result
$(\beta_H,\beta_V)$. In the following, however, we will concentrate 
on the transmission errors induced by the teleportation. As will be
discussed below, these results can be expressed entirely in
terms of the probabilities $P_q(n)$ given by equation (\ref{Pn})
and the well known coherent state fidelity of $(1+q)/2$.

\begin{figure}[ht]
\begin{picture}(350,400)
\put(40,190){\makebox(70,20){\Large $P(q)$}}
\put(40,0){\makebox(300,200){\includegraphics[width=10cm]{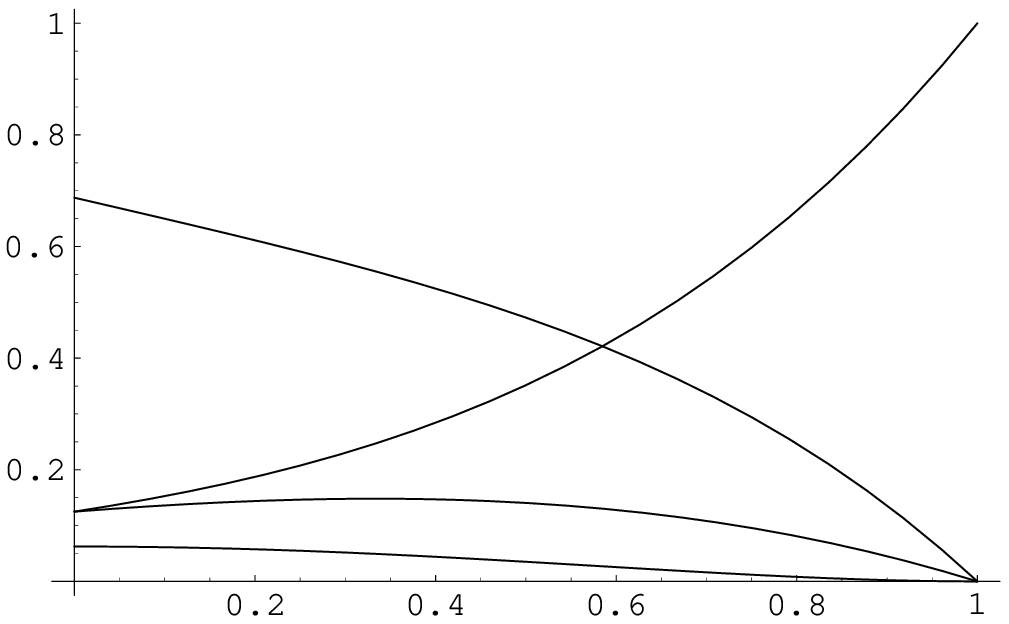}}}
\put(270,100){\makebox(40,10){\Large $P_{\mbox{trans}}(q)$}}
\put(170,35){\makebox(40,10){\Large $P_{\mbox{flip}}(q)$}}
\put(180,50){\makebox(40,10){\Large $P_{\mbox{zero}}(q)$}}
\put(80,140){\makebox(40,10){\Large $P_{n\geq 2}(q)$}}
\put(300,0){\makebox(40,10){\Large $q$}}
\end{picture}
\caption{The $q$ dependence of the probabilities of the polarized photon teleportation.}
\label{pol}
\end{figure}

The total chance of successfully transmitting a photon with the
correct polarization, $P_{\mbox{trans}}$, is equal to the 
product of the probabilities for successfully teleporting a single 
photon, $P_q(1)$, and the probability of successfully teleporting 
the vacuum, $(1+q)/2$, as given by
\begin{eqnarray}
P_{\mbox{trans}}(q)
&=&
\underbrace{\int d^2\beta_H 
 | _H \langle 1 \mid \hat{T}_{Hq}(\beta_H) 
         \mid 1 \rangle_H |^2}_{= \; P_q(1)} 
\underbrace{\int d^2\beta_V 
 | _V \langle 0 \mid \hat{T}_{Vq}(\beta_V) 
         \mid 0 \rangle_V |^2}_{= \; \frac{1+q}{2}}
\nonumber \\
&=& 
\left ( \frac{1+q}{2} \right )^2
\frac{1+q^2}{2} .
\end{eqnarray}
The total chance of a polarization flip while preserving the
total photon number of one, $P_{\mbox{flip}}$, is equal to the 
product of the probabilities for a vacuum output in a one 
photon teleportation, $P_q(0)$, and the reverse situation where 
the vacuum input produces a one photon output. 
Since $\hat{T}_q(\beta)$ is hermitian, however,
these two probabilities are equal and the result reads
\begin{eqnarray}
P_{\mbox{flip}}(q)
&=&
\underbrace{\int d^2\beta_H 
 | _H \langle 0 \mid \hat{T}_{Hq}(\beta_H) 
               \mid 1 \rangle_H |^2}_{=\; P_q(0)}  
\underbrace{\int d^2\beta_V 
 | _V \langle 1 \mid \hat{T}_{Vq}(\beta_V) 
               \mid 0 \rangle_V 
|^2}_{=\; P_q(0)}
\nonumber \\
&=& 
\left ( \frac{1+q}{2} \right )^2
\left ( \frac{1-q}{2} \right )^2 .
\label{pflip}
\end{eqnarray}
Finally, there are also probabilities for changes in the total 
photon number. The chance of obtaining no photon, $P_{\mbox{zero}}$,
is equal to the product of the probabilities for a vacuum output 
in a one photon teleportation, $P_q(0)$, and for the successful 
teleportation of the vacuum given by the coherent state fidelity 
$(1+q)/2$,
\begin{eqnarray}
P_{\mbox{zero}}(q)
&=&
\underbrace{\int d^2\beta_H | _H \langle 0 \mid \hat{T}_{Hq}(\beta_H) 
        \mid 1 \rangle_H |^2}_{=\; P_q(0)} 
\underbrace{\int d^2\beta_V | _V \langle 0 \mid \hat{T}_{Vq}(\beta_V) 
        \mid 0 \rangle_V|^2}_{=\; \frac{1+q}{2}}
\nonumber \\
&=& 
\left ( \frac{1+q}{2} \right )^2
\frac{1-q}{2} .
\end{eqnarray}
The total chance of obtaining more than one photon in the output
can then be obtained by  
\begin{eqnarray}
P_{n\geq 2}(q)
&=&
1-P_{\mbox{flip}}(q)-P_{\mbox{trans}}(q)-P_{\mbox{zero}}(q)
\nonumber \\
&=&
1-\left ( \frac{1+q}{2} \right )^2 \frac{5-4q+3q^2}{4}.
\end{eqnarray}
Fig.(\ref{pol}) shows the $q$ dependence of the above probabilities.
The probability of successful transmission, $P_{\mbox{trans}}$,
increases with increasing entanglement $q$, while the probabilities
of the various error sources decrease. $P_{\mbox{trans}}$ exceeds 
$1/2$ around $q\sim0.7$ and $2/3$ around $q\sim0.8$, illustrating
that the entanglement requirements for high fidelity single photon 
transfers could be fulfilled using the best squeezing sources 
presently available. 
The dominant source of error is the chance of generating additional 
phtons, $P_{n\geq 2}$, while the  
probability of flipping a polarization, $P_{\mbox{flip}}$, 
is always significantly
lower than all the other probabilities. Therefore, the photon loss 
and gain processes are a more serious problem than the flip of a 
polarization for the transmission of the qbit. It is interesting
to compare this new type of error with the post-selection problems
inherent in the previously realized teleportation by entangled
photon pairs \cite{Bou97} using coincidence counting as a trigger. 
In particular, the fidelity of the 
continuous variable teleportation of a polarization could be close
to one if the multi photon outputs could be eliminated by 
post-selection, while the unconditional fidelity at high $q$ values
may proof to be a good alternative in applications where 
post-selection is not an option.

\section{Conclusion}

The properties of continuous variable teleportation of single photon states have been investigated. The difference between the input state and the output state is due to the non-maximal entanglement in the EPR beams shared by Alice and Bob. The field measurement conditions the output state. 
Nearly zero values of the field measurement tend to preserve the initial one photon input state because the eigenstates of the field measurement are close to a photon number states . 
In the intermediate range of field measurement values, photon loss and photon gain processes occur during teleportation. At high values of the field measurement, the probability of photon gain is dominant, corresponding to the high amplitude observed. 
An application of this analysis to the teleportation of a polarized photon shows that the photon loss and gain processes are a more serious problem than the polarization flips. The results imply that unconditional continuous variable teleportation of single photon polarization could be considered an alternative to the post-selected scheme 
\cite{Bou97} using entangled photon pairs.

\end{document}